\documentclass[%
 aip,
rsi,%
 amsmath,amssymb,
 reprint,%
]{revtex4-1}

\usepackage[dvipsnames]{xcolor}
\usepackage{graphicx}
\usepackage{dcolumn}
\usepackage{bm}
\usepackage{lipsum}

\usepackage{dcolumn}
\usepackage{epstopdf}
\usepackage{xcolor}
\usepackage{natbib}
\usepackage{subcaption} 
\usepackage{float}
\usepackage{tabularx}
\usepackage{mathrsfs}
\usepackage{upgreek}
\usepackage[a]{esvect}
\setcitestyle{square}
\usepackage[toc,page]{appendix}
\usepackage[font={small,normalfont}]{caption}
\usepackage[colorlinks=true, allcolors=blue]{hyperref}
\usepackage[normalem]{ulem}
\usepackage[capitalise]{cleveref}

\begin{document}

\title{Electron energy gain due to a laser frequency modulation\\ experienced by electron during betatron motion}

\author{A. Arefiev}
\email[]{aarefiev@ucsd.edu}

\affiliation{Department of Mechanical and Aerospace Engineering, University of California San Diego, La Jolla, CA 92093}

\author{I-L. Yeh}

\affiliation{Department of Physics, University of California San Diego, La Jolla, CA 92093}

\author{L. Willingale}
\affiliation{Center for Ultrafast Optical Science, University of Michigan, Ann Arbor, Michigan 48109}

\date{\today}

\begin{abstract}
Direct laser acceleration of electrons is an important energy deposition mechanism for laser-irradiated plasmas that is particularly effective at relativistic laser intensities in the presence of quasi-static laser-driven plasma electric and magnetic fields. These radial electric and azimuthal magnetic fields provide transverse electron confinement by inducing betatron oscillations of forward-moving electrons undergoing laser acceleration. Electrons are said to experience a betatron resonance when the frequency of betatron oscillations matches the average frequency of the laser field oscillations at the electron position. In this paper, we show that the modulation of the laser frequency caused by the betatron oscillation can be another important mechanism for net energy gain that is qualitatively different from the betatron resonance. Specifically, we show that the frequency modulation experienced by the electron can lead to net energy gain in the regime where the laser field performs three oscillations per betatron oscillation. There is no net energy gain in this regime without the modulation because the energy gain is fully compensated by the energy loss. The modulation slows down the laser oscillation near transverse stopping points, increasing the time interval during which the electron gains energy and making it possible to achieve net energy gain. 
\end{abstract}

\maketitle


\section{Introduction} \label{sec: intro}

In laser-plasma interactions,
direct laser acceleration (DLA) of electrons is a fundamental mechanism for transferring energy from the laser pulse to the plasma. It is  particularly important at relativistic laser intensities, because then the laser is able to generate a forward-moving population of relativistic electrons. These electrons are essential for the development of compact laser-driven gamma-ray \cite{stark.prl.2016,wang.pra.2020,hadjisolomou.pop.2023} and particle sources (e.g. sources of ions~\cite{wilks.pop.2001, dover.prl.2020, gong.prr.2022}, positrons~\cite{Chen_2009_PRL, Chen_2010_PRL}, or neutrons \cite{lancaster.pop.2004, Pomerantz-PhysRevLett.113.184801, gunther.natcomm.2022}).

Electron dynamics during DLA in a laser-irradiated plasma has been extensively studied using experiments~\cite{gahn.prl.1999, willingale.njp.2013, peebles.njp.2017, hussein.njp.2021, rinderknecht.njp.2021}, simulations~\cite{pukhov1999DLA, Robinson_2013_PRL, kemp.pop.2020, jirka.njp.2020}, and analytical theory~\cite{Arefiev_2012_PRL, arefiev2016beyond,khudik.pop.2016, huang.pop.2017,gong.pre.2020, li.prab.2021} with the goal of determining the key factors influencing electron energy gain. There are several aspects that make electron acceleration inside a plasma qualitatively different from electron acceleration in vacuum, which include laser guiding/focusing, superluminal rather than luminal phase velocity, and presence of quasi-static plasma fields. Radial plasma electric and azimuthal plasma magnetic fields that naturally arise during laser propagation through a plasma have attracted particularly attention due to their ability to enhance electron energy gain from the laser. The fields provide transverse electron confinement by deflecting outward moving electrons towards the axis of the laser beam. If the deflections occur on the same time-scale as the oscillations of the laser electric field, then the transverse electron velocity $v_{\perp}$ can remain anti-parallel to the transverse laser electric field $E_{\perp}$ over multiple laser field oscillations. As a result, the electron energy gain can be increased. The extended anti-parallel alignment of $v_{\perp}$ and $E_{\perp}$ is often referred to as a betatron resonance, where the term `betatron' refers to the transverse or betatron oscillations induced by plasma fields. 

The betatron resonance is not the only option for increasing electron energy gain from the transverse laser electric field. During the betatron resonance, the laser electric field performs one oscillation during one betatron oscillation. It was shown in Ref.~[\onlinecite{khudik.pop.2016}] using analytical theory that net energy gain is also possible in the case of three laser oscillations. This phenomenon was named the third-harmonic resonance in Ref.~[\onlinecite{khudik.pop.2016}]. The term itself and the emphasis on the similarity between the third-harmonic resonance and the betatron resonance can perhaps obscure a fundamental difference between the two regimes.

In this paper, we show that the modulation of the laser frequency induced by the betatron motion is the underlying mechanism responsible for net energy gain in the regime where the laser electric field performs three oscillation during one betatron oscillation. In fact, there is no net energy gain in this regime in the absence of the frequency modulation. We employ a test electron model with prescribed laser and plasma fields to elucidate the electron dynamics. We show that the frequency modulation experienced by the electron stretches energy gain regions around stopping points (in the transverse direction) and thus breaks the compensation that occurs without the modulation. In contrast to that, the betatron resonance leads to net energy gain with or without the frequency modulation experienced by the electron. We also show that the period of the frequency modulation experienced by the electron is the reason why no net energy gain occurs for an even number of laser field oscillations.

The rest of the paper consists of seven sections. In \cref{sec: test-model}, we introduce our test-electron model that we then use throughout the paper. In Section~\ref{sec: betatron} and \ref{sec: laser freq}, we provide derivations of the betatron frequency and the frequency of laser field oscillations at the electron location. In Sections~\ref{sec: b-res}, \ref{sec: gain}, and \ref{sec: no gain} we analyze three different regimes with one, three, and two laser field oscillations during a single betatron oscillation. In \cref{sec: summary}, we summarize our key findings.


\section{Test-electron model} \label{sec: test-model}

Electron dynamics during direct laser acceleration can be examined using a test-electron model with prescribed laser and plasma fields. The purpose of this section is to provide a fully self-contained description of the model, as this model is then used to gain insights into the energy gain process. Similar models have been employed in multiple publications on direct laser acceleration~\cite{arefiev.pop.2014, khudik.pop.2016, jirka.njp.2020}.  

Fully self-consistent particle-in-cell (PIC) simulations~\cite{pukhov1996_channel, jansen.ppcf.2018,  gong.pre.2020, wang.pop.2020, jirka.njp.2020, hussein.njp.2021} show that a sufficiently long laser pulse propagating through a plasma generates quasi-static electric and magnetic fields. There is a quasi-static radial electric field caused by expulsion of some electrons out of the laser beam. There is also an azimuthal magnetic field~\cite{stark.prl.2016, jansen.ppcf.2018} sustained by a longitudinal electron current that is generated by the laser during its propagation. In our model, we treat both fields as static. We assume that the azimuthal magnetic field $\bm{B}_{\rm{stat}}$ is generated by a current with uniform current density $j_x=- |j_0|$ and that the radial electric field $\bm{E}_{\rm{stat}}$ is generated by uniform charge density $\rho_{ch} > 0$.

We approximate the laser by a plane monochromatic electromagnetic wave with a phase velocity $v_{ph}$ and frequency $\omega_0$. Under this approximation, we neglect the longitudinal electric field component of the laser. We consider a linearly polarized laser propagating in the positive direction along the $x$-axis, such that
\begin{eqnarray}
     &&\bm{E}_{\rm{plane}} = \bm{e}_y E_0 \cos(\xi + \xi_0), \label{E-laser}\\
     &&\bm{B}_{\rm{plane}} = \bm{e}_z E_{\rm{plane}} / u, \label{B-laser}
\end{eqnarray}
where $E_0$ is the laser amplitude, $c$ is the speed of light, 
\begin{equation}
    u \equiv v_{ph} / c
\end{equation}
is the normalized phase velocity,
\begin{equation} \label{xi}
    \xi = \omega_0 t - \omega_0 x/v_{ph} = \omega_0 t - \omega_0 x/ u c
\end{equation}
is the laser phase, and $\xi_0$ is a phase offset. In general, the laser amplitude $E_0$ varies with $\xi$. We neglect this dependence, so our results are applicable to those regimes where the variation of $E_0$ at electron location is insignificant on the time-scale of the electron energy gain. We introduce a normalized laser amplitude
\begin{equation} \label{eq: a0}
    a_0 = |e|E_0/m_e c \omega_0,
\end{equation}
where $m_e$ is the electron mass. We also introduce 
the vacuum laser wavelength $\lambda_0 = 2 \pi c / \omega_0$ that we use for normalization.

In this study, we limit our analysis to flat particle trajectories in the $(x,y)$-plane ($z = 0$ and $p_z = 0$). Non-planar electron trajectories can become unstable~\cite{ArefievPOP2016non-planar}, so they are not considered in this work. We therefore only need the knowledge of $\bm{B}_{\rm{stat}}$ and $\bm{E}_{\rm{stat}}$ in the $(x,y)$-plane. Under our assumptions, the radial electric field in the $(x,y)$-plane is
\begin{equation}
    \bm{E}_{\rm{stat}} = 2 \kappa \frac{y}{\lambda_0^2} \frac{m_e c^2}{|e|} \bm{e}_y  \label{E_stat}
\end{equation}
where 
\begin{equation}
    \kappa \equiv \pi^2 \omega_{ch}^2 / \omega_0^2
\end{equation}
and 
\begin{equation}
    \omega_{ch}^2 \equiv 4 \pi |e| \rho_{ch}/m_e
\end{equation}
is a plasma frequency calculated using $\rho_{ch}$. The plasma magnetic field is
\begin{equation}
    \bm{B}_{\rm{stat}} = \frac{m_e c^2}{|e|} \left[ \nabla \times \bm{a}_{\rm{stat}} \right],
\end{equation}
where 
\begin{equation}
    \bm{a}_{\rm{stat}} = \bm{e}_x  \alpha (y^2+z^2)/\lambda_0^2.
\end{equation}
The dimensionless parameter 
\begin{equation}
    \alpha \equiv \pi \lambda_0^2 |j_0|/J_A.
\end{equation}
represents the ratio of the current flowing through a circular area with radius $\lambda_0$ to the classical Alfv\'en current $J_A=m_e c^3/|e|$. The field in the $(x,y)$-plane is 
\begin{equation}
    \bm{B}_{\rm{stat}} = - \frac{m_e c^2}{|e|} \frac{2 \alpha y}{\lambda_0^2} \bm{e}_z. \label{B_stat}
\end{equation}

The electron dynamics is then described by the following system of coupled differential equations:
\begin{eqnarray}
&& \frac{d \bm{p}}{d t} = - |e| \bm{E} - \frac{|e|}{\gamma m_e c} \left[ \bm{p} \times \bm{B} \right], \label{dpdt} \\
&& \frac{d \bm{r}}{d t} = \frac{c}{\gamma} \frac{\bm{p}}{m_e c}, \label{drdt} 
\end{eqnarray}
where $\bm{E} = \bm{E}_{\rm{plane}} + \bm{E}_{\rm{stat}}$, $\bm{B} = \bm{B}_{\rm{plane}} + \bm{B}_{\rm{stat}}$, and
\begin{equation} \label{gamma}
    \gamma = \sqrt{1 + \bm{p}^2/ m_e^2 c^2}
\end{equation}
is the relativistic factor of the electron. We assume that $a_0$ is sufficiently low to neglect the effect of radiation reaction~\cite{gong.scirep.2019,yeh.njp.2021,jirka.njp.2020}. Our test-electron model consists of Eqs.~(\ref{dpdt}) and (\ref{drdt}) where the fields are the fields of the laser given by Eqs.~(\ref{E-laser}) and (\ref{B-laser}) and the fields of the plasma given by Eqs.~(\ref{E_stat}) and (\ref{B_stat}). In addition to initial conditions, the model requires four input parameters: $a_0$, $\kappa$, $\alpha$, and  $u$.


\section{Betatron oscillations} \label{sec: betatron}

The plasma electric and magnetic fields are responsible for transverse electron confinement.  An electron with a transverse momentum stays near the axis while performing transverse oscillations that we will refer to as betatron oscillations. The oscillations typically take place in conjunction with longitudinal electron motion in the direction of laser propagation.

We start by deriving a conservation law for a general case where the electron experiences the fields of the plasma and the fields of the laser. The work done on the electron per unit time is a scalar product of Eq.~(\ref{dpdt}) with $\bm{v} = \bm{p}/\gamma m_e$. We take into account the specific dependence of $\bm{E}_{\rm{stat}}$ on $y$ given by Eq.~(\ref{E_stat}) to find from the resulting equation that
\begin{equation} \label{eq: 16}
    \frac{d}{dt} \left( \gamma + \kappa \frac{y^2}{\lambda_0^2} \right) = - \frac{|e| E_{\rm{plane}}}{m_e c} \frac{p_y}{\gamma m_e c}.
\end{equation}
\Cref{eq: 16} quantifies the energy exchange between the electron and the electric fields of the laser and plasma. These fields change the transverse momentum $p_y$, whereas the magnetic fields redistribute the energy between the transverse and longitudinal motion. It is convenient to re-write the longitudinal component of Eq.~(\ref{dpdt}) using the expression for $\bm{B}_{\rm{stat}}$ given by Eq.~(\ref{B_stat}) as
\begin{equation} \label{eq: 17}
    \frac{d}{dt} \left( - \frac{p_x}{m_e c} + \alpha \frac{y^2}{\lambda_0^2} \right) =  \frac{|e| B_{\rm{plane}}}{m_e c} \frac{p_y}{\gamma m_e c}.
\end{equation}
This equation quantifies changes in $p_x$ due to the electron motion in the magnetic fields of the plasma and laser. We multiply both sides of \cref{eq: 17} by $u$ and add the resulting equation to \cref{eq: 16}. After taking into account that $B_{\rm{plane}} = E_{\rm{plane}}/u$, the terms on the right-hand side cancel each other out and we obtain
\begin{equation} \label{eq: 18}
    \frac{d}{dt} \left( \gamma - u \frac{p_x}{m_e c} + \left[ u \alpha  + \kappa \right]  \frac{y^2}{\lambda_0^2} \right) = 0.
\end{equation}
The derived equation shows that 
\begin{equation} \label{S_exp}
   S =  \gamma - u \frac{p_x}{m_e c} + \left[ u \alpha  + \kappa \right]  \frac{y^2}{\lambda_0^2}
\end{equation}
is conserved during electron betatron oscillations. This is a well-known result in DLA research.

It might appear odd that the derived conservation law involves $u$ without involving the laser fields. Indeed, in the limit of $E_{\rm{plane}} \rightarrow 0$ and $B_{\rm{plane}} \rightarrow 0$, $S$ still involves the laser phase velocity even though there is no longer any laser field acting on the electron. In order to understand what is happening at $E_{\rm{plane}} \rightarrow 0$ and $B_{\rm{plane}} \rightarrow 0$, we need to go back to Eqs.~(\ref{eq: 16}) and (\ref{eq: 17}). In this limit, these two equations turn into 
\begin{equation} \label{eq: 16-2}
    \frac{d}{dt} \left( \gamma + \kappa \frac{y^2}{\lambda_0^2} \right) = 0
\end{equation}
and
\begin{equation} \label{eq: 17-2}
    \frac{d}{dt} \left( - \frac{p_x}{m_e c} + \alpha \frac{y^2}{\lambda_0^2} \right) = 0,
\end{equation}
which means that, in the absence of the laser, we have two conservation laws rather than one. At $E_{\rm{plane}} = 0$ and $B_{\rm{plane}} = 0$, Eq.~(\ref{eq: 18}) is simply a sum of these two equations without providing any additional information, with the terms that involve $u$ canceling each other and the terms without $u$ canceling each other as well.

\begin{figure}[htb]
    \begin{center}
    \includegraphics[width=1\columnwidth,clip]{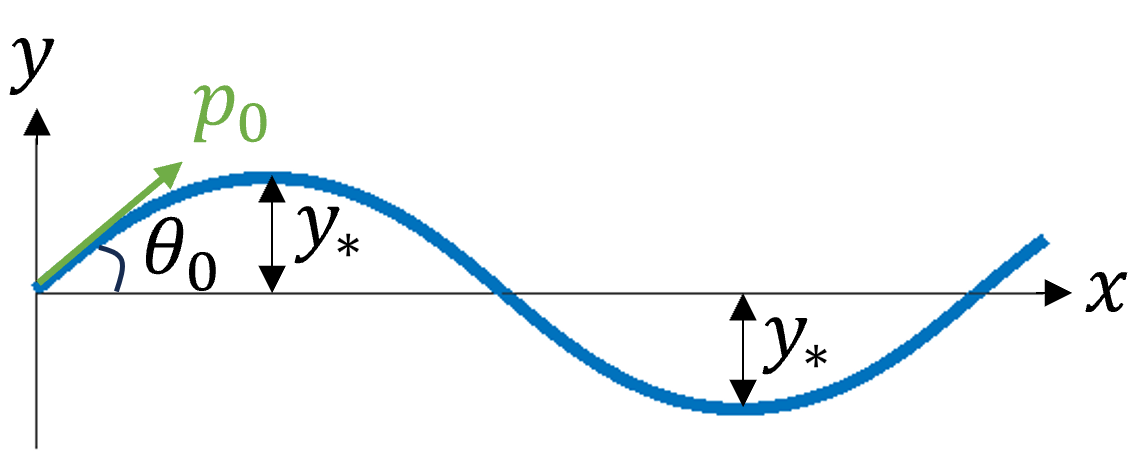}
    \caption{\label{fig:ystar_theta0} Betatron oscillation of an ultra-relativistic electron moving to the right along the $x$-axis.  
} 
    \end{center}
\end{figure}

We now consider an ultra-relativistic electron that performs betatron oscillations in the absence of the laser, as shown in Fig.~\ref{fig:ystar_theta0}. We focus on a scenario where the electron crosses the $x$-axis at a small angle $|\theta_0| \ll 1$. To find the amplitude of betatron oscillations $y_*$, we note that the maximum transverse displacement is achieved at a turning point where $p_y = 0$. Equations~(\ref{eq: 16-2}) and (\ref{eq: 17-2}) at $y = y_*$ give a system of equations that determines $y_*$ in terms of electron parameters at $y = 0$:
\begin{eqnarray}
    && \gamma + \kappa \frac{y_*^2}{\lambda_0^2} = \gamma_0, \\
    &&  - \frac{p_x}{m_e c} + \alpha \frac{y_*^2}{\lambda_0^2} = - \frac{p_0 \cos \theta_0}{m_e c},
\end{eqnarray}
where $\gamma_0$ and $p_0$ are the electron $\gamma$-factor and the total momentum at $y = 0$. This system has a simple approximate solution for $|\theta_0| \gg 1/\gamma_0$. We add the two equations, which yields
\begin{equation} \label{eq: 25}
    \left( \kappa + \alpha \right) \frac{y_*^2}{\lambda_0^2} = \left( \gamma_0 - \frac{p_0 \cos \theta_0}{m_e c} \right) - \left[ \gamma - \frac{p_x}{m_e c} \right].
\end{equation}
The expression in the round brackets on the right hand side of \cref{eq: 25} can be approximated by
\begin{eqnarray}
    \gamma_0 - \frac{p_0 \cos \theta_0}{m_e c} &=& \gamma_0 - \frac{p_0}{m_e c} + \frac{p_0}{m_e c} \left( 1 - \cos \theta_0 \right) \nonumber \\
    & \approx &  \frac{1}{2 \gamma_0} + \gamma_0 \frac{\theta_0^2}{2} \approx \gamma_0 \theta_0^2 / 2,  \label{round bracket terms}
\end{eqnarray}
where it is taken into account that the electron is relativistic and that $\theta_0 \gg 1/\gamma_0$. The expression inside the square brackets in \cref{eq: 25} can be roughly estimated as $1/2 \gamma_0$, where we disregard the change in $\gamma$. This quantity is much smaller than $\gamma_0 \theta_0^2 / 2$. We thus neglect the terms inside the square brackets in \cref{eq: 25}
and obtain the following expression for the amplitude of betatron oscillations:
\begin{equation} \label{eq: y_max no laser}
    \frac{y_*}{\lambda_0} \approx \left( \frac{\gamma_0 \theta_0^2}{2 (\kappa + \alpha)} \right)^{1/2}.
\end{equation}

The change in electron kinetic energy during a betatron oscillation is relatively small for $|\theta_0| \ll 1$. We find directly from \cref{eq: 16-2} that the maximum change in $\gamma$ is given by
\begin{equation}
    \Delta \gamma = |\gamma - \gamma_0| = \kappa y_*^2 / \lambda_0^2.
\end{equation}
Using the expression for $y_*$ given by \cref{eq: y_max no laser}, we find that
\begin{equation} \label{delta gamma}
    \frac{\Delta \gamma}{\gamma_0} = \frac{\kappa \theta_0^2}{2 (\kappa + \alpha)} < \theta_0^2.
\end{equation}
For $|\theta_0| \ll 1$, we have $\Delta \gamma \ll \gamma_0$, so that the relative energy change is indeed small. The underlying reason for the small change can be understood by recognizing that it is only the transverse plasma electric field that can reduce the electron $\gamma$. An electron with $|\theta_0| \ll 1$ has most of the energy associated with the longitudinal rather than transverse motion when it crosses the axis. It therefore does not have a lot of energy that it can give up when it starts to lose $p_y$ during its outward motion. 

We conclude this section by deriving an expression for the frequency of betatron oscillations. We describe betatron oscillations using a new variable $\psi$ that we call the betatron phase, with
\begin{equation} \label{y beta 0}
    y = y_* \sin \psi.
\end{equation}
We define the frequency of betatron oscillations as
\begin{equation} \label{def: omega beta}
    \omega_{\beta} \equiv d \psi / dt.
\end{equation} 

To obtain an equation for $\omega_{\beta}$, we start by adding \cref{eq: 16-2} to \cref{eq: 17-2}. After integrating the resulting equation, we obtain
\begin{equation} \label{eq: main 0}
     \gamma - \frac{p_x}{m_e c} + (\kappa + \alpha) \frac{y^2}{\lambda_0^2} = \gamma_0 - \frac{p_0 \cos \theta_0}{m_e c} .
\end{equation}
It follows from the definition of $\gamma$ given by \cref{gamma} that
\begin{equation} \label{eq: 32}
    \gamma - \frac{p_x}{m_e c} \approx \frac{p_y^2}{2 p_x m_e c} \approx \frac{\gamma}{2} \frac{v_y^2}{c^2},
\end{equation}
where we took into account that $v_y/c = p_y / \gamma m_e c$ and then replaced $p_x$ with $\gamma m_e c$. We differentiate \cref{y beta 0} with respect to $t$ to find that
\begin{equation} \label{vy beta}
    v_y = y_* \frac{d \psi}{dt} \cos \psi = y_* \omega_{\beta} \cos \psi.
\end{equation}
As a result, we have
\begin{equation} \label{eq:26}
    \gamma - \frac{p_x}{m_e c} \approx \frac{\gamma}{2} \frac{y_*^2 \omega_{\beta}^2}{c^2} \cos^2 \psi.
\end{equation}
The next step is to substitute this expression and the expression for $y$ given by \cref{y beta 0} into \cref{eq: main 0}, which yields: 
\begin{equation} 
     \frac{\gamma}{2} \frac{y_*^2 \omega_{\beta}^2}{c^2} \cos^2 \psi + (\kappa + \alpha) \frac{y_*^2}{\lambda_0^2} \sin^2 \psi = \gamma_0 - \frac{p_0 \cos \theta_0}{m_e c} .
\end{equation}
After taking into account that the right-hand side is approximately given by Eq.~(\ref{round bracket terms}) and using the expression for $y_*$ given by \cref{eq: y_max no laser}, we obtain
\begin{equation} \label{eq: 36}
     \left( \frac{\gamma}{2} \frac{\omega_{\beta}^2}{c^2} \frac{\lambda_0^2}{\kappa + \alpha} \right)\cos^2 \psi + \sin^2 \psi = 1 .
\end{equation}
\Cref{eq: 36} must be satisfied for any $\psi$, which requires the combination inside the brackets to be equal to unity. This condition provides an expression for the frequency of betatron oscillations:
\begin{equation} \label{eq: omega beta 0}
    \frac{\omega_{\beta}}{\omega_0} =  \frac{1}{\sqrt{2} \pi} \left[ \frac{\kappa + \alpha}{\gamma} \right]^{1/2},
\end{equation}
where it is taken into account that $\lambda_0 = 2 \pi c / \omega_0$. 


\section{Frequency of laser field oscillations experienced by electron} \label{sec: laser freq}

In the presence of a laser, the forward moving electron we are considering sees the laser fields oscillate with frequency
\begin{equation} \label{omega prime 0}
    \omega' \equiv d \xi / dt,
\end{equation}
where $\xi$ is the laser phase at the electron location. This frequency is much lower than $\omega_0$ because of the forward relativistic motion. More importantly, the betatron oscillations strongly modulate $\omega'$ by changing the longitudinal velocity of the electron. 

Our goal in this section is to derive an expression for $\omega'$ in terms of the betatron phase $\psi$. This requires re-deriving the expression for the amplitude of betatron oscillations $y_*$ while taking into account that the electron can exchange energy with the laser. We again consider a forward-moving electron with $p_x \gg |p_y| \gg m_e c$. We focus on a regime where the superluminosity is weak ($\delta u \ll 1$), treating $\delta u$ as a small parameter. Our derivation leverages the fact that the quantity $S$ given by Eq.~(\ref{S_exp}) remains constant during the electron motion. We substitute $u = 1 + \delta u$ into the second term on the right-hand side of Eq.~(\ref{S_exp}) to obtain the following approximate expression that we will use:
\begin{equation}
    S  \approx \left( \gamma - \frac{p_x}{m_e c} \right) -  \gamma \delta u  +  [u \alpha + \kappa] \frac{y^2}{\lambda_0^2}.
    \label{S_exp-3}
\end{equation}
Here we took into account that the difference between $\gamma$ and $p_x/m_e c$ is relatively small. We neglect this difference in the $\delta u$-term by replacing $p_x/m_e c$ with $\gamma$. 

We use the conservation of $S$ to find the amplitude of betatron oscillations. We consider a single betatron oscillation and write down the expression for $S$ at a turning point, $y = y_*$. We neglect the contribution from the expression in the round brackets in Eq.~(\ref{S_exp-3}) to obtain
\begin{equation} \label{temp-1}
    S  \approx - \gamma \delta u +  [u \alpha + \kappa] \frac{y_*^2}{\lambda_0^2}.
\end{equation}
As discussed in Sec.~\ref{sec: betatron}, the contribution from the neglected terms is relatively small compared to the $y_*^2$-term if $\theta_0 \gg 1/\gamma_0$, where $\gamma_0$ and $\theta_0$ are the electron relativistic factor and the angle between $\bm{p}$ and the $x$-axis at $y = 0$. We find directly from Eq.~(\ref{temp-1}) that
\begin{equation} \label{ymax}
    \frac{y_*}{\lambda_0} \approx \left[ \frac{S + \gamma \delta u}{\kappa + u \alpha} \right]^{1/2} \approx \left[ \frac{S + \gamma \delta u}{\kappa + \alpha} \right]^{1/2},
\end{equation}
where we set $u \approx 1$ in the denominator due to the smallness of $\delta u$. The derived expression is particularly meaningful when the change in $\gamma$ is relatively small over one betatron oscillation. The convenience of Eq.~(\ref{ymax}) is that it enables us to track the evolution of $y_*$ over many betatron oscillations as the electron energy changes due to the interaction with the laser. In the case of $\delta u=0$, the amplitude of betatron oscillations remains unchanged. In the superluminal case ($\delta u > 0$), the behavior of $y_*$ is qualitatively different. The amplitude starts to grow with the increase of electron energy once the relativistic factor reaches the level at which $\gamma \gg S/\delta u$.

It is worth pointing out that, over a single betatron oscillation, Eq.~(\ref{ymax}) is approximately the same as Eq.~(\ref{eq: y_max no laser}) for small $\delta u$. We first make a comparison for a case without the laser. We take into account that 
\begin{eqnarray}
    S &=& \gamma_0 - u \frac{p_0}{m_e c} \cos \theta_0 \approx \frac{p_0}{m_e c} - u \frac{p_0}{m_e c}\left( 1 - \frac{\theta_0^2}{2} \right) \nonumber \\
    & \approx & -\delta u \gamma_0 + \gamma_0 \theta_0^2/ 2
\end{eqnarray}
and find that
\begin{equation}
    S + \gamma \delta u \approx \gamma_0 \theta_0^2/ 2 + (\gamma - \gamma_0) \delta u.
\end{equation}
In the absence of the laser, $|\gamma - \gamma_0| \leq \Delta \gamma$, where $\Delta \gamma$ is the quantity given by \cref{delta gamma}. The amplitude of the last term on the right-hand side can then be estimated as $\delta u \gamma_0 \theta_0^2$, which shows that this term is negligible compared to the first term on the right-hand side. After neglecting the last term, \cref{ymax} becomes identical to \cref{eq: y_max no laser}. In the presence of the laser, the same argument holds if the second term remains negligible.

We are interested in a regime where the accelerating electron is able to continue gaining energy from the laser over many betatron oscillations. In this case, the change in $\gamma$ and thus the change in $y_*$ in a single betatron oscillation are relatively small. We can then again describe betatron oscillations as
\begin{equation} \label{y beta}
    y \approx y_* \sin \psi,
\end{equation}
where $y_*$ is assumed to be independent of $\psi$. 

To derive an expression for $\omega'$, we start with the definition given by Eq.~(\ref{omega prime 0}) that we re-write using the expressions for $\xi$ and $v_x$ given by Eqs.~(\ref{xi}) and (\ref{drdt}):
\begin{eqnarray} \label{omega prime orig}
    \omega' &=& \omega_0 \left( 1 - \frac{v_x}{v_{ph}} \right) = \frac{\omega_0}{\gamma} \left( \gamma - \frac{1}{u} \frac{p_x}{m_e c}  \right).  
\end{eqnarray}
After substituting $u = 1 + \delta u$ and retaining only the linear $\delta u$-term, we obtain
\begin{eqnarray}
    \omega' & \approx & \frac{\omega_0}{\gamma} \left( \gamma - \frac{p_x}{m_e c} + \gamma \delta u \right).  \label{omega prime}
\end{eqnarray}
In the $\delta u$-term we replaced $p_x/m_e c$ with $\gamma$ neglecting the small difference between the two. The next step is to obtain an expression for $\gamma - p_x/m_e c$ from Eq.~(\ref{S_exp-3}). We substitute $y= y_* \sin \psi$ and use the expression for $y_*$ given by \cref{ymax} to find that
\begin{equation} \label{eq:gamma -  px}
    \gamma - \frac{p_x}{m_e c} =  \left( S + \gamma \delta u \right) \cos^2 \psi.
\end{equation}
After substituting this expression into Eq.~(\ref{omega prime}), we obtain an expression for $\omega'$ in terms of the electron $\gamma$-factor and the betatron phase $\psi$:
\begin{equation} \label{omega prime v3 new}
    \frac{\omega'}{\omega_0} = \frac{1}{\gamma} \left[ (S + \gamma \delta u) \cos^2 \psi + \gamma \delta u \right].
\end{equation}

During a single betatron oscillation ($-\pi \leq \psi < \pi$), $\omega'$ performs two oscillations. It has two maxima that correspond to $y = 0$: one at $\psi = -\pi$ and one at $\psi = 0$. It also has two minima that correspond to $v_y = 0$: one at $\psi = -\pi/2$ and one at $\psi = \pi/2$. The modulation of $\omega'$ is a direct consequence of the fact that $\omega' \propto 1 - v_x/v_{ph}$ [see Eq.~(\ref{omega prime orig})]. In the course of a betatron oscillation, the velocity amplitude $v$ remains almost unchanged, whereas the velocity vector $\bm{v}$ experiences an oscillation with respect to the $x$-axis during which $v_y$ changes its sign. The longitudinal velocity $v_x$ remains positive and this is why $v_x$ performs two oscillations during a single betatron oscillation. In order to make the discussed dependence more apparent, we re-write \cref{omega prime v3 new} as
\begin{equation} \label{omega prime v4}
    \frac{\omega'}{\omega_0} = \frac{1}{2 \gamma} \left[ (S + 3 \gamma \delta u) + (S + \gamma \delta u) \cos (2 \psi) \right].
\end{equation}


To conclude this section, we show that the expression for $\omega_{\beta}$ given by \cref{eq: omega beta 0} is still valid in the presence of the laser. In this section, we used the conservation of $S$ to derive the expression for $\gamma - p_x/m_e c$ given by Eq.~(\ref{eq:gamma -  px}). This expression should agree with the expression given by \cref{eq:26} that was derived using the relation $y = y_* \sin \psi$. The two expressions are consistent if
\begin{equation} \label{eq:27}
    \omega_{\beta} = \left[ \frac{2}{\gamma} ( S + \delta u \gamma ) \right]^{1/2} \frac{c}{y_*}.
\end{equation}
To obtain the final expression, 
we substitute the definition of $y_*$ given by \cref{ymax} into \cref{eq:27} and express $\lambda_0$ in terms of $\omega_0$. The result reads  
\begin{equation} \label{eq: omega beta}
    \frac{\omega_{\beta}}{\omega_0} =  \frac{1}{\sqrt{2} \pi} \left[ \frac{\kappa + \alpha}{\gamma} \right]^{1/2}.
\end{equation}
This result matches the betatron frequency derived in the absence of the laser.


\section{Betatron resonance} \label{sec: b-res}

In the test-electron model of \cref{sec: test-model}, the electron gains kinetic energy only from the transverse electric field that consists of the static plasma field and the oscillating field of the laser. The electron continuously exchanges energy with the static field of the plasma, but there is no net energy gain over one betatron oscillation. The laser field is time-dependent and, therefore, a net energy gain over one oscillation becomes possible.

We calculate the energy gain over one betatron oscillation by integrating \cref{eq: 16} over $t$. We first express $p_y$ in terms of $v_y$ using the relation $p_y/\gamma m_e = v_y$. We then
use the expression for $E_{\rm{plane}}$ given by Eq.~(\ref{E-laser}), the expression for $v_y$ given by \cref{vy beta}, and the definition of $a_0$ given by \cref{eq: a0} to re-write the right-hand side of \cref{eq: 16} as
\begin{equation}
    - \frac{|e| E_{\rm{plane}}}{m_e c} \frac{p_y}{\gamma m_e c} = - a_0 \omega_0 \cos(\xi + \xi_0) \frac{y_* \omega_{\beta}}{c} \cos(\psi).
\end{equation}
We assume that the relative energy gain over one betatron oscillation is small, which allows us to treat $y_*$ as constant. It is convenient to re-write the time integral as an integral over the betatron phase $\psi$ using the definition of $\omega_{\beta}$ given by \cref{eq: omega beta}, which yields
\begin{equation} \label{eq: 16 int 2}
    \Delta \gamma = - a_0 \frac{y_* \omega_0}{c} \int_{-\pi}^{\pi} \cos(\xi + \xi_0) \cos(\psi) d \psi.
\end{equation}
To compute the integral in \cref{eq: 16 int 2}, we need to know the wave phase $\xi$ as a function of the betatron phase $\psi$. We find the corresponding dependence from \cref{omega prime 0}. We multiply both sides of \cref{omega prime 0} by $dt$ and integrate the resulting equation. After replacing the time integration with an integral over $\psi$, we obtain
\begin{equation} \label{eq: xi}
    \xi = \int_0^{\psi} \frac{\omega'}{\omega_{\beta}} d \psi.
\end{equation}
Without any loss of generality, we set $\xi = 0$ at $\psi = 0$. 


In contrast to the betatron frequency, the frequency of the laser oscillations perceived by the electron $(\omega')$ is modulated. Its average over one betatron oscillation is 
\begin{equation}
    \langle \omega' \rangle = \frac{1}{2 \pi} \int_{- \pi}^{\pi} \omega' d \psi.
\end{equation}
Using the expression given by \cref{omega prime v4}, we find that
\begin{equation} \label{omega prime av}
    \frac{\langle \omega' \rangle}{\omega_0} = \frac{1}{2\gamma} \left[ S + 3 \gamma \delta u \right].
\end{equation}    
We now re-write the expression for $\xi$ as
\begin{equation} \label{eq: xi v2}
    \xi = \frac{\langle \omega' \rangle}{\omega_{\beta}} \psi + \int_0^{\psi} \frac{\omega' - \langle \omega' \rangle}{\omega_{\beta}} d \psi.
\end{equation}
We substitute $\omega'$ from \cref{omega prime v4} and perform the integration to obtain 
\begin{equation} \label{xi general}
    \xi = \frac{\langle \omega' \rangle}{\omega_{\beta}} \psi + \frac{\omega_0}{\omega_{\beta}} \frac{S + \gamma \delta u}{4 \gamma} \sin(2 \psi).
\end{equation}
The second term on the right-hand side results from the modulation of $\omega'$ caused by the betatron oscillations.

\begin{figure}[htb]
    \begin{center}
    \includegraphics[width=1\columnwidth,clip]{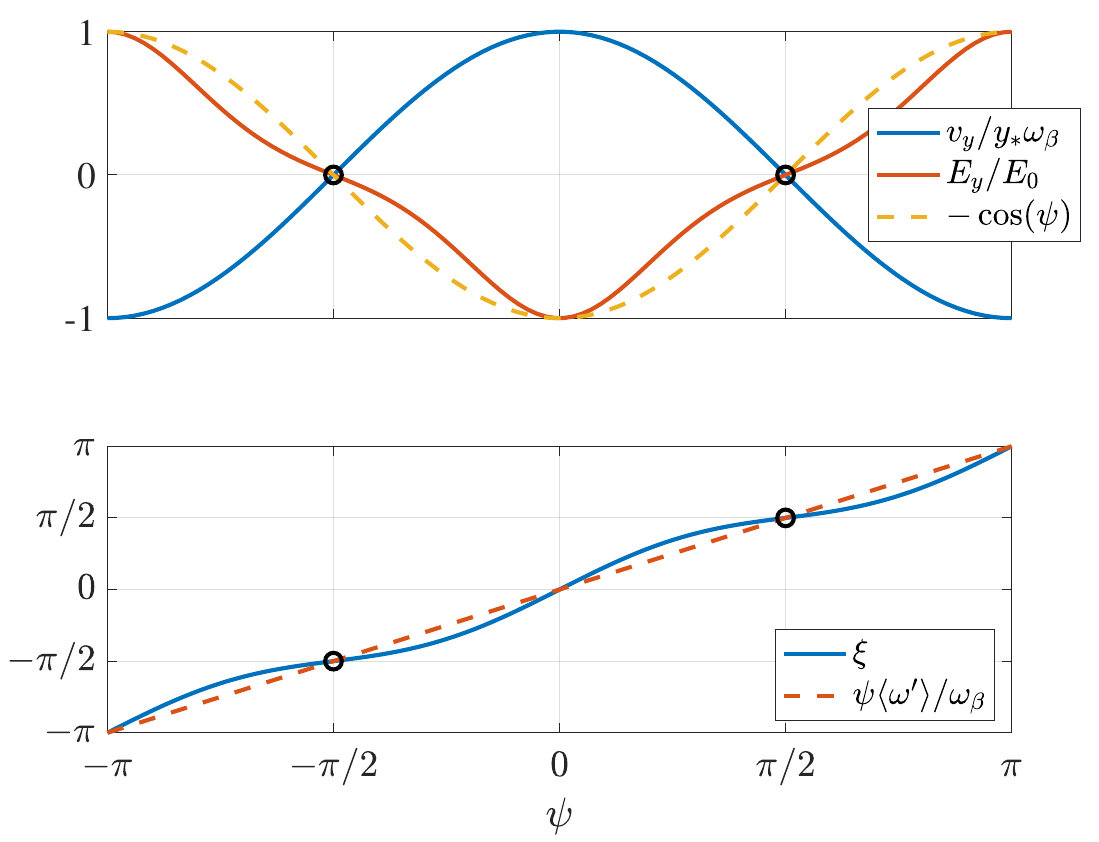}
    \caption{Betatron resonance with $\langle \omega' \rangle = \omega_{\beta}$ and $\xi_0 = -\pi$. The amplitude of the frequency modulation defined by \cref{eq: C} is set at $C_1 = 0.3$. The circles mark turning points in the transverse direction.}
    \label{fig:betatron_resonance}
    \end{center}
\end{figure}

One should expect net energy gain for $\langle \omega' \rangle = \omega_{\beta}$, because the average frequency of the laser oscillations experienced by the electron matches the frequency of betatron oscillations. Indeed, if we disregard the modulation of $\xi$, we readily find from \cref{eq: 16 int 2} that 
\begin{equation}
    \Delta \gamma = - \cos(\xi_0) \Delta_{\max},    
\end{equation}
where
\begin{equation} \label{Delta_max}
    \Delta_{\max} = \pi a_0 y_* \omega_0 / c.
\end{equation}
The highest gain is achieved for $\xi_0 = -\pi$, because then the laser field remains anti-parallel to the electron velocity over the entire betatron oscillation. This is the so-called betatron resonance.

The modulation of $\omega'$ caused by the betatron oscillation is detrimental to the energy gain during the betatron resonance. To find the energy gain without neglecting the modulation, we first re-write \cref{xi general} for a specific case of $\langle \omega' \rangle = \ell \omega_{\beta}$, where $\ell$ is an integer. Since $\langle \omega' \rangle$ is given by \cref{omega prime av}, the condition $\langle \omega' \rangle = \ell \omega_{\beta}$ requires a specific value of the betatron frequency:
\begin{equation} \label{omega beta new}
    \frac{\omega_{\beta}}{\omega_0} = \frac{S + 3 \gamma \delta u}{2 \ell \gamma}.
\end{equation}   
Taking this expression into account, we find from \cref{xi general} that
\begin{equation} \label{xi-simple}
    \xi = \ell \psi + C_{\ell} \sin ( 2 \psi),
\end{equation}
where
\begin{equation} \label{eq: C}
    C_{\ell} =  \frac{\ell}{2} \frac{S + \gamma \delta u}{S + 3 \gamma \delta u}.
\end{equation}
We next substitute the expression for $\xi$ with $\ell$ set to unity ($\ell = 1$) into \cref{eq: 16 int 2} and use the Hansen-Bessel formula, i.e. the integral Bessel function representation, to find that
\begin{equation} \label{delta}
    \Delta \gamma = \Delta_{\max} \left[ J_0 (C_1) - J_1 (C_1) \right] = \Delta_{\max} \left[ 1 - \delta_{mod}  \right],
\end{equation}
where $J_0$ and $J_1$ are the Bessel functions of the first kind and
\begin{equation}
    \delta_{mod} = 1 + J_1 (C_1) - J_0 (C_1)
\end{equation}
is the relative reduction in the energy gain due to the modulation of $\omega'$. The modulation amplitude given by $C_1$ reduces with the increase of $\gamma$ and $\delta u$. It has the highest value of $C_1 = 1/2$ in the luminal regime ($\delta u = 0$). It reaches its lowest value of $C_1 = 1/6$ in the limit of $\gamma \rightarrow \infty$. The corresponding range for the relative reduction in the energy gain is $0.30 \geq \delta_{mod} \geq 0.09$.

\Cref{fig:betatron_resonance} illustrates a betatron resonance with $C_1 = 0.3$. The upper panel shows the electron velocity $v_y$ and laser field $E_y$ as functions of the betatron phase $\psi$ during a single betatron oscillation. The lower panel of \cref{fig:betatron_resonance} shows how the laser phase $\xi$ increases with $\psi$ during the same betatron oscillation. The modulation of $\omega'$ modulates $d \xi/ d \psi$, but the net change of the laser phase $\xi$ between the turning points $\psi = \pm \pi/2$ (shown with circles) remains equal to $\pi$. As a result, the laser field remains anti-parallel to $v_y$ during the entire betatron oscillation despite the frequency modulation.

\begin{figure}[htb]
    \begin{center}
    \includegraphics[width=0.7\columnwidth,clip]{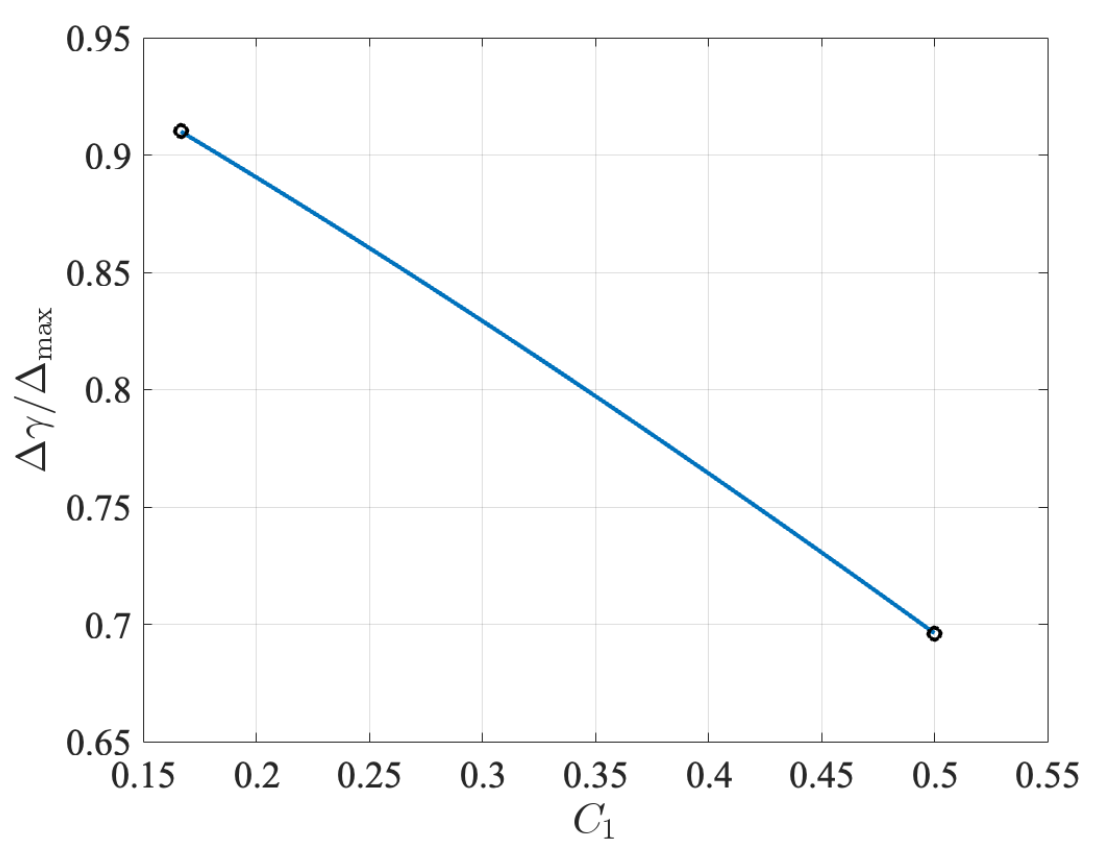}
    \caption{Electron energy gain for $\langle \omega' \rangle = \omega_{\beta}$ and $\xi_0 = -\pi$. The constant $C_1$ is defined by \cref{eq: C} with $\ell = 1$. The circles mark the boundaries of the possible range, $1/6 \leq C_1 \leq 1/2$.}
    \label{fig:C1}
    \end{center}
\end{figure}

The modulation of $\omega'$
reduces the energy gain by inducing the modulation of $\xi$ that is visible in the lower panel of \cref{fig:betatron_resonance}. Indeed, $E_y$ in the upper panel of \cref{fig:betatron_resonance} looks markedly different from $\cos (\psi)$: the low field regions around $\psi = \pm \pi/2$ are extended and strong field regions around $\psi = 0$ and $\psi = \pm \pi$ are reduced. Note that the betatron phase $\psi$ increases linearly with time $t$. Therefore, we conclude that the modulation of $\xi$ reduces the time the electron spends interacting with the strongest field while also increasing the time the electron spends interacting with $|E_y| \ll E_0$. 



To explain the origin of the modulation of $\xi$, we take a closer look at the electron dynamics. The points with $\psi = \pm \pi/2$ are the transverse turning points. At the turning points, $v_y$ goes to zero and $v_x$ reaches its highest value. As a result, $d \xi / d\psi \propto d \xi / dt \propto 1 - v_x / v_{ph}$ reaches its lowest value. We find from \cref{xi-simple} that $d \xi / d \psi = 1 - 2 C_1$ at $\psi = \pm \pi/2$. The points with $\psi = 0$ and $\psi = \pm \pi$ are the points where the electron crosses the axis. Here $|v_y|$ reaches its highest value and $v_x$ reaches its lowest value. As a result, $d \xi / d\psi \propto d \xi / dt \propto 1 - v_x / v_{ph}$ reaches its highest value. We find from \cref{xi-simple} that $d \xi / d \psi = 1 + 2 C_1$ at $\psi = 0$ and $\psi = \pm \pi$.

\Cref{fig:C1} summarizes our findings by showing the dependence of the energy gain over a single betatron oscillation on the frequency modulation amplitude $C_1$ for $\xi_0 = -\pi$. The circles mark the boundaries of the possible range, $1/6 \leq C_1 \leq 1/2$. According to \cref{eq: C}, $C_1$ decreases with the increase of $\gamma$, which means that electron energy gain reduces the amplitude of the modulation. We can therefore conclude from \cref{fig:C1} that electron energy gain becomes more efficient (more energy per betatron oscillation) with the increase of electron energy.


\section{Energy gain at $\langle \omega' \rangle = 3\omega_{\beta}$ due to experienced frequency modulation} \label{sec: gain}

We have already shown that an electron experiences a betatron resonance if the laser field performs one full oscillation during a betatron oscillation. This condition is equivalent to $\langle \omega' \rangle = \omega_{\beta}$. If the laser field performs two or more full oscillations, then no net energy gain is possible in the absence of $\omega'$ modulations. Indeed, it follows directly from \cref{eq: 16 int 2} that $\Delta \gamma = 0$
for $\xi = \psi \langle \omega' \rangle / \omega_{\beta}$, where $\langle \omega' \rangle = \ell \omega_{\beta}$ and $\ell$ is an integer greater than one ($\ell > 1$). In what follows, we show that the $\omega'$ modulation induced by the betatron motion leads to net energy gain for $\ell = 3$.

\begin{figure}[htb]
    \begin{center}
    \includegraphics[width=0.7\columnwidth,clip]{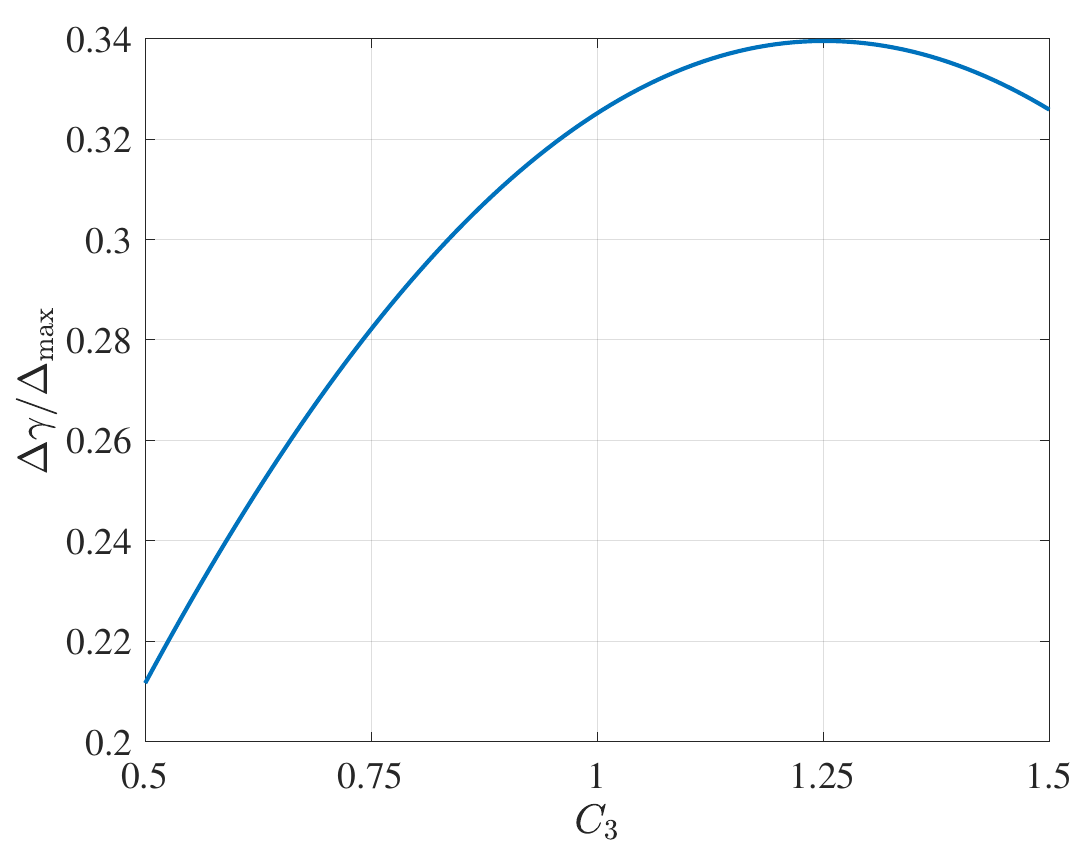}
    \caption{Electron energy gain for $\langle \omega' \rangle = 3\omega_{\beta}$ and $\xi_0 = 0$. The constant $C_3$ is defined by \cref{eq: C} with $\ell = 3$.}
    \label{fig:C3}
    \end{center}
\end{figure}

To find the energy gain due to the frequency modulation, we substitute the expression for $\xi$ given by \cref{xi-simple} into \cref{eq: 16 int 2} and set $\ell = 3$. We use the Hansen-Bessel formula, i.e. the integral Bessel function representation, to find that, for a given phase offset $\xi_0$, the energy gain over one betatron oscillation is
\begin{equation} \label{delta-3}
    \Delta \gamma = \cos(\xi_0) \Delta_{\max} \left[ J_1 (C_3) - J_2 (C_3) \right],
\end{equation}
where $\Delta_{\max}$ is defined by \cref{Delta_max}. According to \cref{eq: C}, $C_3$ varies between 0.5 and 1.5, reaching its highest value in the luminal regime ($\delta u = 0$). The expression inside the square brackets in \cref{delta-3} remains positive for this range of $C_3$, which means that the highest energy gain for a given value of $C_3$ occurs for the case with no phase offset, i.e. $\xi_0 = 0$.

\Cref{fig:C3} shows how the energy gain given by \cref{delta-3} depends on $C_3$ for $\xi_0 = 0$. In the luminal regime ($\delta u = 0$), the value of $C_3$ is fixed ($C_3 = 1.5$). However, in the superluminal regime ($\delta u > 0$), $C_3$ decreases with the increase of $\gamma$. As a consequence, the energy gain becomes less efficient after $C_3$ exceeds 1.25 due to the increase of $\gamma$ ($C_3 \approx 1.25$ is the location of the maximum). The energy gain per betatron oscillation is the lowest for $\gamma \delta u \gg S$, because, in this limit, $C_3$ approaches its lowest value of 0.5. The discussed behavior of $\Delta \gamma$ differs from that given by \cref{delta} for $\ell = 1$. In the case of the betatron resonance, the energy gain becomes more efficient with the increase of $\gamma$ because of the reduced frequency modulation amplitude, as seen in \cref{fig:C1}.

\begin{figure}[t]
    \begin{center}
    \includegraphics[width=\columnwidth,clip]{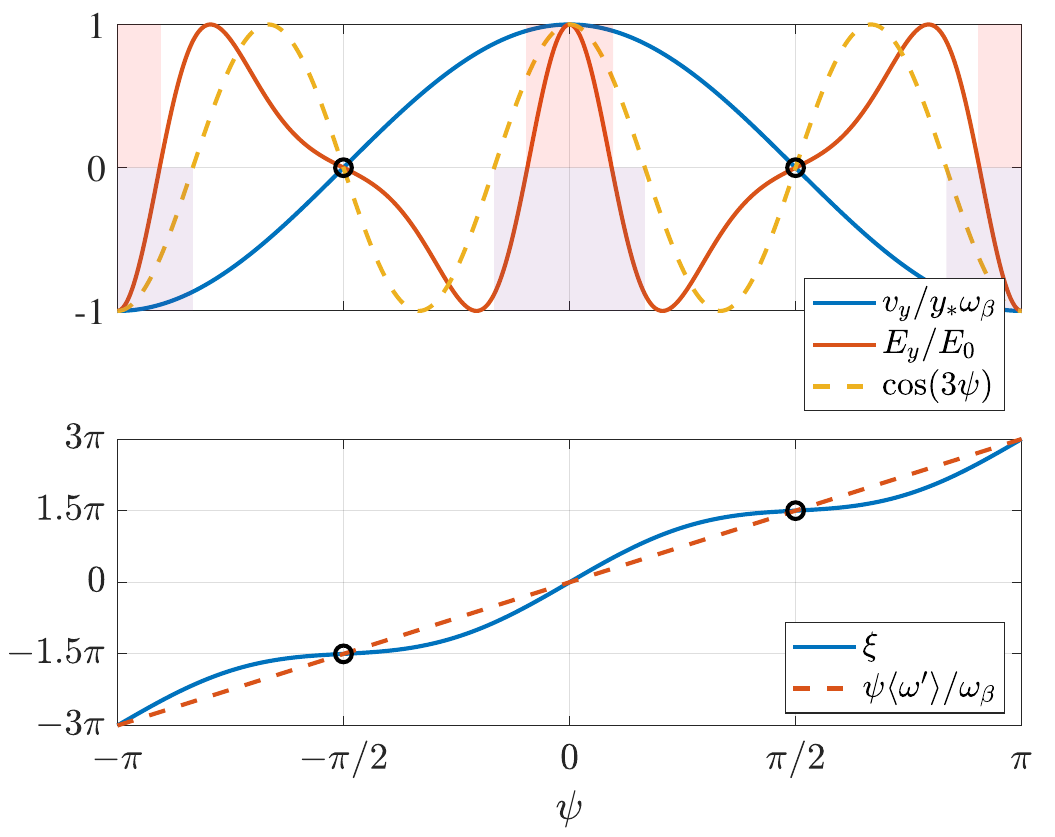}
    \caption{Electron dynamics and field evolution at $\langle \omega' \rangle = 3\omega_{\beta}$ and $\xi_0 = 0$. The amplitude of the frequency modulation defined by \cref{eq: C} is set at $C_3 = 1.25$. The circles mark turning points in the transverse direction. The upper shaded regions show the range of $\psi$ where $v_y$ is parallel to $E_y$. The lower shaded regions show the range of $\psi$ where $v_y$ is parallel to $\cos(3 \psi)$.}
    \label{fig:third_order}
    \end{center}
\end{figure}

\Cref{fig:third_order} shows the electron dynamics and field evolution for $C_3 = 1.25$ and $\xi_0 = 0$. The net change of the laser phase $\xi$ between the turning points $\psi = \pm \pi/2$ (shown with circles) is $3\pi$. This is also the case when there is no frequency modulation, as indicated with the dashed curve in the upper panel of \cref{fig:third_order}. In both cases, $v_y$ is parallel to $E_y$ near $\psi = 0$ and $\psi = \pm \pi$. This is the part of the trajectory where the electron loses energy (upper shaded regions in the upper panel of \cref{fig:third_order}). On the other hand, $v_y$ is anti-parallel to $E_y$ near the turning points. This is the part of the trajectory where the electron gains energy (these regions are not shaded in the upper part of the upper panel of \cref{fig:third_order}). In the absence of the modulation, the gain is fully compensated by the energy loss. For reference, the lower shaded regions show the range of $\psi$ where $v_y$ is parallel to $\cos(3 \psi)$ and thus the electron loses energy in the absence of the modulation ($\cos(3 \psi)$ represents the field without the modulation of $\omega'$). The modulation of $\omega'$ extends the range of $\psi$ where $v_y$ is anti-parallel to $E_y$ and reduces the range of $\psi$ where $v_y$ is parallel to $E_y$ (the shaded regions become narrower), which causes net energy gain over one betatron oscillation. 

\begin{figure}[t]
    \begin{center}
    \includegraphics[width=\columnwidth,clip]{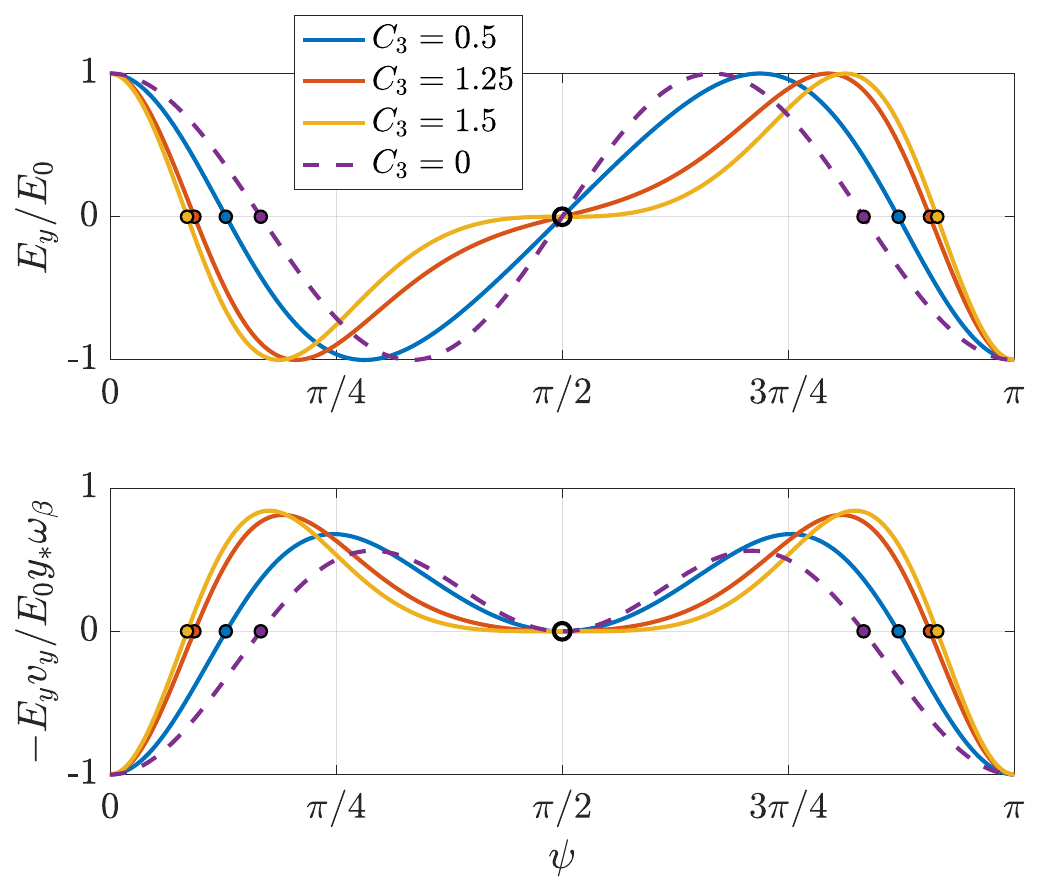}
    \caption{Impact of the frequency modulation on the evolution of the laser electric field for $\langle \omega' \rangle = 3\omega_{\beta}$ and $\xi_0 = 0$. The solid curves in the upper panel show $E_y / E_0$ for three different values of $C_3$. The dashed curve is the field profile without the frequency modulation ($C_3 = 0$). The lower panel shows the normalized value of $-E_y v_y$ versus $\psi$ for the same field profiles. The empty circle marks a transverse turning point ($v_y = 0$). The filled circles show where $E_y v_y$ changes its sign.}
    \label{fig:E_vs_C3}
    \end{center}
\end{figure}

The upper panel in \cref{fig:E_vs_C3} illustrates the impact of the frequency modulation induced by the betatron motion on the evolution of the transverse laser electric field $E_y$. The amplitude of the modulation is quantified by the parameter $C_3$ that can vary between 0.5 and 1.5. As a reference, the dashed curve shows the case without the modulation ($C_3 = 0$). In this case, there is no net energy gain or, equivalently, the 
integral of $E_y v_y$ over $\psi$ goes to zero. The lower panel in \cref{fig:E_vs_C3} shows the variation of $E_y v_y$ over one half of the betatron oscillation (the other half looks identical). The filled circles indicate where $E_y v_y$ changes its sign. The energy gain occurs between the two filled circles. It is fully compensated by the energy loss outside of this interval.

The role of the experienced frequency modulation can be understood by tracking the location of the filled circles in the lower panel of \cref{fig:E_vs_C3}. These circles indicate where $E_y v_y$ changes its sign. At $C_3 = 0.5$, the circles are further apart than in the case without the modulation ($C_3 = 0$). This is because $d \xi / d \psi$ is lowered near the turning point located at $\psi = \pi / 2$.  As a result, the region with $-E_y v_y > 0$ is wider and the electron gains more energy than it loses. An increase of $C_3$ from 0.5 to 1.25 pushes the filled circles even further apart, extending the region with $-E_y v_y > 0$ and increasing the energy gain. Increasing $C_3$ above 1.25 is counterproductive, because the width of the region with $-E_y v_y > 0$ increases insignificantly. At the same time, the derivative of $E_y/E_0$ near the turning point becomes noticeably smaller. This stretches the region of low values of $E_y/E_0$ and thus reduces the energy gain.



\section{Lack of net energy in the presence of the frequency modulation at $\langle \omega' \rangle = 2\omega_{\beta}$} \label{sec: no gain}

In \cref{sec: gain}, we showed that the frequency modulation caused by the betatron motion leads to net energy gain for $\langle \omega' \rangle = 3\omega_{\beta}$. In this section, we examine the case with $\langle \omega' \rangle = 2\omega_{\beta}$ to show that net energy gain requires for the number of laser field oscillations during a betatron oscillation to be odd.

\begin{figure}[htb]
    \begin{center}
    \includegraphics[width=\columnwidth,clip]{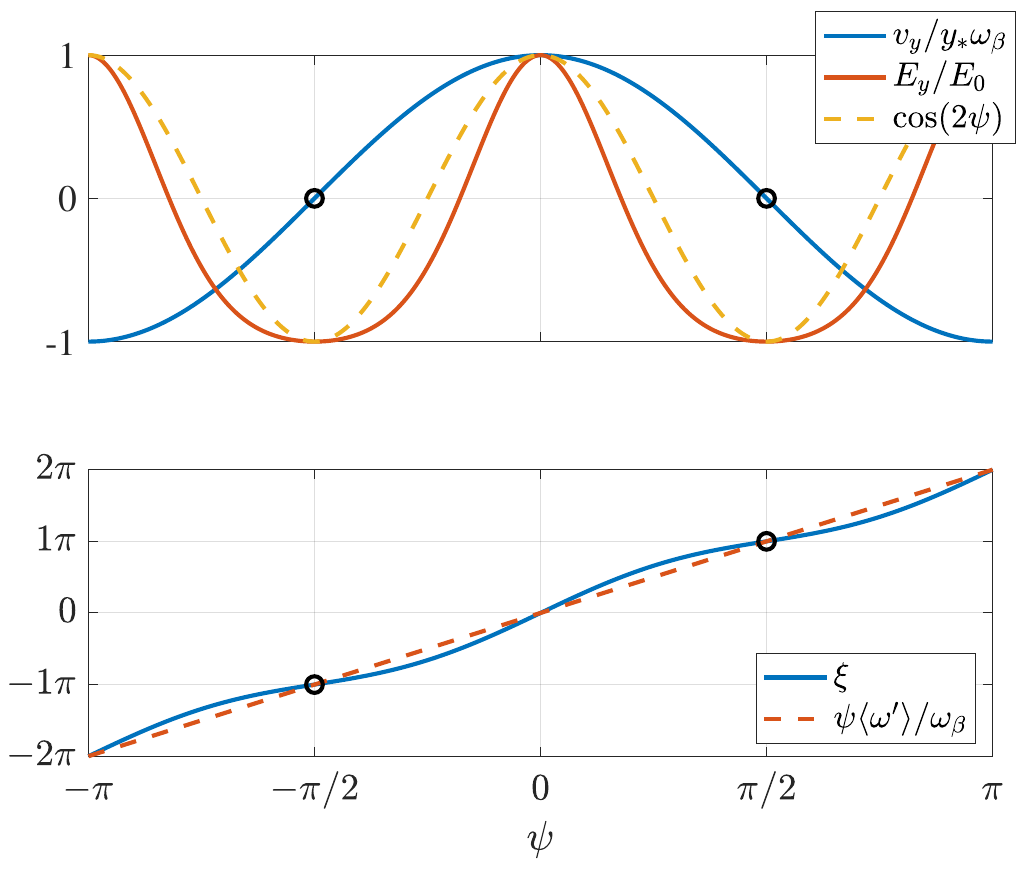}
    \caption{Electron dynamics and field evolution at $\langle \omega' \rangle = 2\omega_{\beta}$ and $\xi_0 = 0$. The amplitude of the frequency modulation defined by \cref{eq: C} is set at $C_2 = 0.5$. The circles mark turning points in the transverse direction.}
    \label{fig:second_order}
    \end{center}
\end{figure}

\Cref{fig:second_order} shows the evolution of the laser field $E_y$ and the phase variable $\xi$ for $\langle \omega' \rangle = 2\omega_{\beta}$ and $\xi_0 = 0$. A fundamental difference with the regimes where $\langle \omega' \rangle = \omega_{\beta}$ or $\langle \omega' \rangle = 3\omega_{\beta}$ is that $E_y$ no longer changes its sign at the turning points. Instead, the turning points are the locations of the minima of $E_y$. This behavior remains unaltered by the frequency modulation. 

\begin{figure}[t]
    \begin{center}
    \includegraphics[width=\columnwidth,clip]{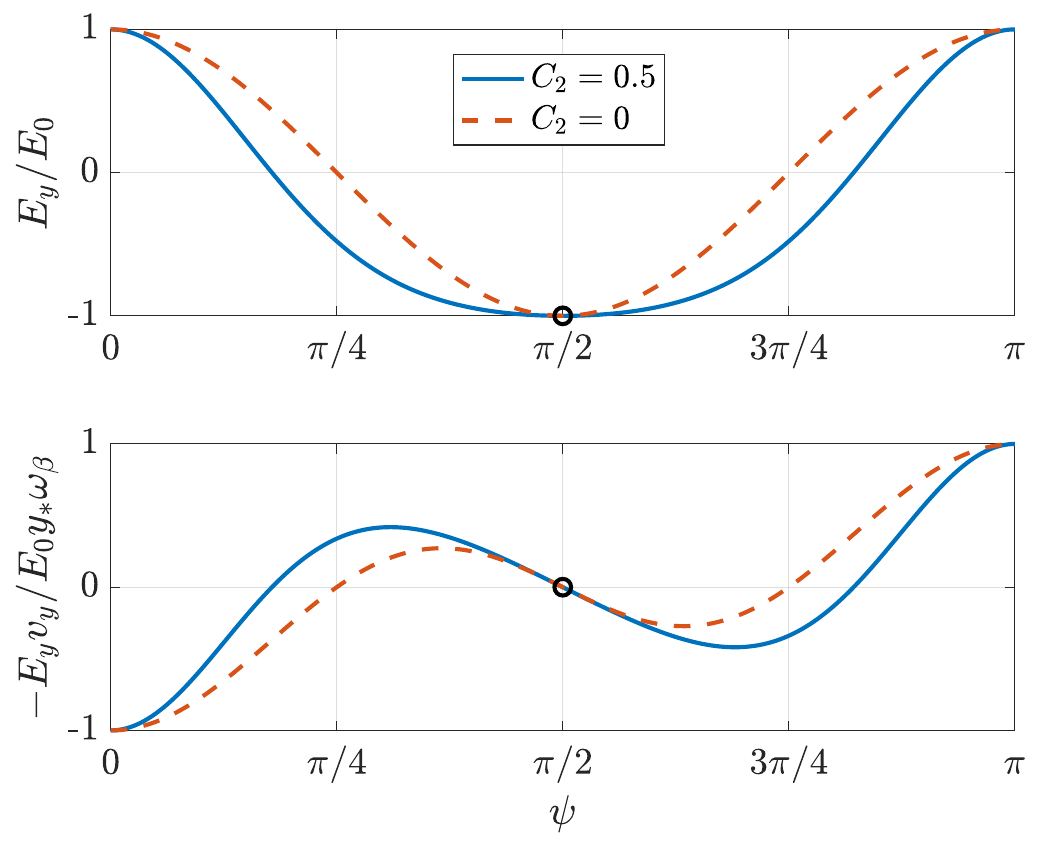}
    \caption{Impact of the frequency modulation on the evolution of the laser electric field for $\langle \omega' \rangle = 2\omega_{\beta}$ and $\xi_0 = 0$. The solid curve in the upper panel shows $E_y / E_0$ for $C_2 = 0.5$. The dashed curve is the field profile without the frequency modulation ($C_2 = 0$). The lower panel shows the normalized value of $-E_y v_y$ versus $\psi$ for the same field profiles. The empty circle marks a transverse turning point ($v_y = 0$). }
    \label{fig:E_vs_C2}
    \end{center}
\end{figure}

As seen in the upper panel of \cref{fig:E_vs_C2}, the frequency modulation elongates the region with a strong $E_y$ near the turning point (marked with a circle). The elongation is symmetrical with respect to $\psi = \pi/2$, so that $E_y (-\psi') = E_y (\psi')$, where $\psi' = \psi - \pi/2$. Since the velocity changes its sign at $\psi = \pi/2$, we have $v_y(-\psi') E_y (-\psi') = -v_y(\psi) E_y (\psi)$. This means that the energy loss at $\psi' < 0 $ cancels out the energy gain at $\psi' >0$. As a result, the elongation caused by the frequency modulation produces no net energy gain in the considered regime.  

The obtained result is not specific to the choice of the phase offset $\psi_0$. In fact, there is no net energy gain at $\langle \omega' \rangle = 2\omega_{\beta}$ for any $\xi_0$. To show this, we rewrite the expression for $\Delta \gamma$ given by \cref{eq: 16 int 2} as
\begin{eqnarray}
    \Delta \gamma &=& - \frac{\Delta_{\max}}{\pi}  \int_{-\pi}^{\pi} \cos(\xi + \xi_0) \cos(\psi) d \psi \nonumber \\
    &=& \frac{\Delta_{\max}}{\pi} \sin(\xi_0) \int_{-\pi}^{\pi} \sin(\xi) \cos(\psi) d \psi \nonumber \\
    &-& \frac{\Delta_{\max}}{\pi} \cos(\xi_0) \int_{-\pi}^{\pi} \cos(\xi) \cos(\psi) d \psi. \label{eq: 67}
\end{eqnarray}
It follows from the expression for $\xi$ given by \cref{xi general} that, in general, 
\begin{equation}
    \xi (-\psi) = -\xi(\psi).
\end{equation}
Therefore, the integral that involves the product $\sin(\xi) \cos(\psi)$ is equal to zero and we are left only with the integral that contains $\cos(\xi) \cos(\psi)$:
\begin{equation} \label{eq: 69}
        \Delta \gamma =
    - \frac{\Delta_{\max}}{\pi} \cos(\xi_0) \int_{-\pi}^{\pi} \cos(\xi) \cos(\psi) d \psi.
\end{equation}
It is important to point out that the value of the integral in \cref{eq: 69} is independent of $\xi_0$. To provide a specific example, we have already considered the case with $\xi_0 = 0$ and showed that the integral is equal to zero. By changing $\xi_0$, we can only change the multiplier, whereas the integral will remain equal to zero. Therefore, $\Delta \gamma$ remains equal to zero for all possible values of $\xi_0$. 

The period of the frequency modulation is the primary reason why there is no net energy gain in the considered regime. As seen in \cref{omega prime v4}, $\omega'$ has two minima during each betatron oscillation. These minima correspond to two transverse turning points. If the number of laser field oscillations is three, then these minima can be matched with the locations where $E_y = 0$. In this case, the electron gains energy near the turning points and the frequency modulation emphasizes the energy gain while reducing energy losses. The same logic applies to the regimes with $\ell>3$ where $\ell$ is odd. If the number of laser field oscillations is two, then it is not possible to match up the turning points with zeroes of $E_y$, which is necessary for taking advantage of the frequency modulation. As a result, there is no net energy gain despite the frequency modulation. The same logic applies to the regimes with $\ell>2$ where $\ell$ is even.

\section{Summary} \label{sec: summary}

In this paper, we examined the role of the laser frequency modulation caused by the betatron oscillation of the electron in quasi-static plasma electric and magnetic fields. It is shown that the frequency modulation experienced by the electron reduces the net energy gain per betatron oscillation during the betatron resonance. However, there is a regime where the presence of the modulation is beneficial. In this regime, the laser performs three oscillations per betatron period. There is no energy gain without the modulation, because the energy gain is fully compensated by the energy loss. The experienced frequency modulation slows down the laser field oscillation near transverse stopping points, which then prolongs the time interval during which the electron experiences energy gain. As a result, a net energy gain becomes possible. The same mechanism applies to other regimes where the number of laser oscillations per betatron period is odd and greater than one. If the number of oscillations is even, then the energy gain remains fully compensated by the energy loss despite the frequency modulation.

The results for an odd number of laser oscillations $\ell$ over a single betatron oscillation can be summarized with a single expression. For a given odd $\ell$, the change of the relativistic $\gamma$ factor over one betatron oscillation is
\begin{equation} \label{delta-gamma-general}
    \frac{\left(\Delta \gamma\right)_{\ell}}{\Delta_{\max}} = (-1)^{k+1} \cos(\xi_0)  \left[ J_k (C_{\ell}) - J_{k+1} (C_{\ell}) \right],
\end{equation}
where 
\begin{equation}
    k = (\ell - 1)/2.
\end{equation}
This expression can be verified by performing the integral in \cref{eq: 69} with $\xi$ given by \cref{xi general}. \Cref{delta-gamma-general} provides a general expression that is valid for any odd $\ell$ rather than just for $\ell = 1$ and $\ell = 3$. For even values of $\ell$, we have $\left(\Delta \gamma\right)_{\ell} = 0$, i.e. there is no net energy gain.


\section*{Acknowledgments}

This material is based upon work supported by the Department of Energy National Nuclear Security Administration under Award Number DE-NA0004030. Simulations were performed with EPOCH (developed under UK EPSRC Grants No. EP/G054940/1, No. EP/G055165/1, and No. EP/G056803/1) using HPC resources provided by TACC at the University of Texas at Austin. 

\section*{References}
\bibliography{Collection}


\end{document}